# Violation of a Bell inequality in two-dimensional spin-orbit hypoentangled subspaces

**Lixiang Chen,** * **Xiancong Lu, and Zhongqun Cheng**

*Department of Physics, Xiamen University, Xiamen 361005, China*

*Corresponding author: chenlx@xmu.edu.cn

Based on spin-orbit coupling induced by *q*-plates, we present a feasible experimental proposal for preparing two-dimensional spatially inhomogeneous polarizations of light. We further investigate the quantum correlations between these inhomogeneous polarizations of photon pairs generated by spontaneous parametric down-conversion, which in essence describe the so-called *hypoentanglement* that is established between composite spin-orbit variables of photons. The violation of the Clauser-Horne-Shimony-Holt-Bell inequality is predicted with $S = 2\sqrt{2}$ to illustrate the entangled nature of the cylindrical symmetry of spatially inhomogeneous polarizations. © 2011 Optical Society of America

## 1. INTRODUCTION

Entanglement is the quintessential feature of quantum mechanics and is at the heart of the Einstein-Podolsky-Rosen (EPR) paradox [1]. Photon pairs generated by spontaneous parametric downconversion (SPDC) have been a reliable resource for testing the violation of Bell's inequality, which proves that local realism leads to algebraic predictions contradicted by quantum mechanics [1]. To access violation of the original Bell inequality, it is essential to base on a Hilbert space of dimension two [2]. Besides, entanglement is a fundamental concept that is indifferent to the specific physical realization of that space. Polarization formed much of the early work, which is in essence the entanglement of spin angular momentum of light [3]. Besides, violation of a Bell inequality has been carried out in *every* degree of freedom of spin, orbital angular momentum (OAM) and energy-time in a *hyperentangled* state [4]. In analogy with polarization, various two-dimensional OAM subspaces [5] and optical vortex links [6] were also constructed to violate the Bell inequality using spatial light modulators. Recently, *hybrid* entanglement, established between two *different* degrees of freedom of a single photon [7, 8] or of a photon pair [9-13], was again shown to access the violation of a Bell inequality.

Different from *hyperentanglement* or *hybrid* entanglement, *hypoentanglement* also represents a new resource useful for quantum information processing [14]. A hypoentanglement can be realized based on the *composite* energy-time-spin-orbit or frequency-polarization variables of photon pairs [15, 16]. In those schemes, the production of hypoentanglement however requires a high interferometric stability [15, 16]. Besides, to our best knowledge, no schemes have been proposed to characterize the hypoentanglement. To this end, we here propose a feasible experimental scheme to test the violation of a Bell inequality for two-dimensional spin-orbit hypoentanglement, benefiting from the novel spin-orbit coupling induced by *q*-plates. Of particular interest is that our results further reveal the entangled nature between the cylindrical polarization symmetries of SPDC photon pairs.

## 2. SPATIALLY INHOMOGENEOUS POLARIZATIONS FROM Q-PLATES

We prepare the spatially inhomogeneous polarizations with cylindrical symmetry based on *q*-plates recently devised by Marrucci and coworkers [17, 18]. The *q*-plates are a novel type of inhomogeneous anisotropic waveplates having a singular transverse pattern of the birefringent optical axis, which is described in the polar coordinate by, $\alpha(\phi) = q\phi$, where *q* denotes the topological singularity charge. Thus, the spin-orbit coupling effect induced by such *q*-plates on an incoming photon with an OAM of $m\hbar$ can be represented by an operator [7],

$$\hat{Q}(q) = |R, m+2q\rangle\langle L, m| + |L, m-2q\rangle\langle R, m|, \qquad (1)$$

where $|L\rangle$ and $|R\rangle$ are eigenstates of photon spin, namely, left- and right-handed circular polarizations, respectively. The *q*-plates have variety of fascinating applications [19], including optimal quantum cloning of OAM qubits [20], quantum information transfer form spin to OAM subspace [21], hybrid entanglement swapping [22] and open-destination teleportation [23].

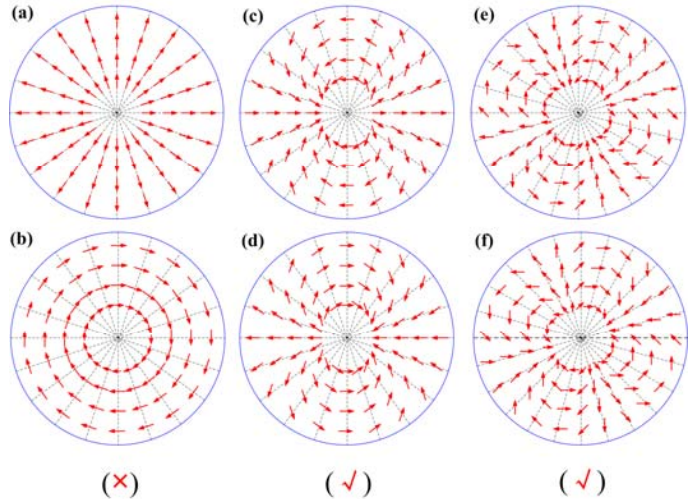

Fig. 1 (Color online) Spatial distribution of instantaneous electric field of photons emerging from *q*-plates: (a) *q*=1/2, θ=0; (b) *q*=1/2, θ=π/2; (c) *q*=1, θ=0; (d) *q* =-1, θ=0; (e) *q*=3/2, θ=π/4; (f) *q*=-3/2, θ=-π/4. (√) or (×) denotes quantum correlation or not in individual column.

Here, we focus on another application of the generation and manipulation of spatially inhomogeneous polarizations with cylindrical symmetry. Assume that the incoming planer photon of $m=0$ is linearly polarized at angle $\theta$, namely, $|\psi\rangle_{in} = \cos\theta|H\rangle + \sin\theta|V\rangle = (e^{-i\theta}|L\rangle + e^{i\theta}|R\rangle)/\sqrt{2}$, interaction with $q$-plates described by Eq. (1) then modifies the outgoing photon state,

$$|\psi\rangle_{out} = \frac{1}{\sqrt{2}}(e^{-i\theta}|R,+2q\rangle + e^{i\theta}|L,-2q\rangle). \tag{2}$$

In the frame of Jones space, we can rewrite Eq. (2) in the simple form of a two-dimensional vector,

$$\begin{pmatrix} E_x \\ E_y \end{pmatrix} = \begin{pmatrix} \cos(2q\phi - \theta) \\ \sin(2q\phi - \theta) \end{pmatrix}. \tag{3}$$

The mathematical interest of Eq. (3) is the revelation of the cylindrical polarization symmetry possessed by the outgoing photons. Along polar angle $\phi$, linear polarization points at angle $2q\phi - \theta$. For integer or half-integer $q$-plate, the polarization evolution with $\phi$ is therefore continuous and periodic. Obviously, the cylindrical symmetry can be tuned by adopting suitable $q$-plates (via $q$) or by changing the incident polarization (via $\theta$). Figure 1(a) and 1(b) illustrate two basic examples of radial and azimuthal polarizations, where $q=1/2$, $\theta=0$ and $\theta=\pi/2$ are used, respectively. One can see that the local orientation of linear polarization at $\phi$ can be also modulated by $\theta$ for a specific $q$-plate. Besides, the cycle of cylindrical polarization symmetry is determined by $T = \pi/|q|$. We illustrate this point in Fig. 1(c)-1(f) for comparison. Figure 1(c) and 1(d) exhibit a two-fold rotational symmetry by using $q$-plates of $q=\pm 1$, while in Fig. 1(e) and 1(f), the rotational symmetry is three-fold with $q = \pm 3/2$. Recently there has been an increasing interest in light beams with spatially inhomogeneous polarizations, as they can expand the functionality of optical systems [24]. However, our main purpose here is to reveal the entangled nature of these spatially inhomogeneous polarizations of hypoentangled photon pairs generated by SPDC.

## 3. VIOLATION OF A BELL INEQUALITY FOR SPIN-ORBIT HYPOENTANGLEMENT

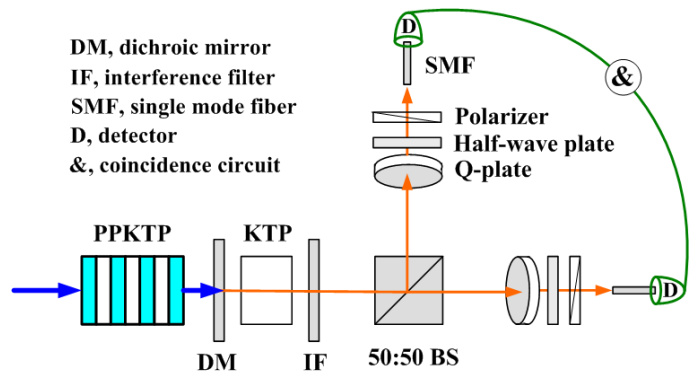

Fig. 2 (Color online) The proposed experimental setup.

The proposed scheme is sketched in Fig. 2. A periodically poled potassium titanyl phosphate (PPKTP) as down-converter enables a high-efficiency down-conversion by quasi-phase matching [25]. The 397nm pump beam with horizontal polarization is weakly focused to a 10mm PPKTP with a grating period of $8.84 \mu m$, where the type-II SPDC is done. The orthogonally polarized 795nm signal and idler photons are generated collinearly and separated from the pump ones by a dichroic mirror (DM). Their timing information is erased by a following KTP compensator. After 795nm filter (IF), they are further steered to a 50:50 non-polarizing beam splitter (BS). Within a coincidence window, the postselection of one photon in transmitted path (*t*) and the other in reflected (*r*) path thus produces the polarization entangled state, $|\Psi\rangle_{spin} = (|H\rangle_t|V\rangle_r + |V\rangle_t|H\rangle_r)/\sqrt{2}$. Besides, owing to angular momentum conservation, the down-converted photons also exhibit high-dimensional OAM entangled state [26, 27], $|\Psi\rangle_{orbit} = \sum_m C_m |m\rangle_t |-m\rangle_r$, where $C_m$ is related to the OAM entangled spectrum. Thus, this setup can be utilized as a spin-orbit hyper-entangled source,

$$|\Psi\rangle_{hyper} = |\Psi\rangle_{spin} \otimes |\Psi\rangle_{orbit}. \quad (4)$$

Based on spin-orbit hyper-entanglement, it has been demonstrated how to beat the channel capacity limit [28], increase Shannon dimensionality [29] and remotely prepare single-photon hybrid entanglement [30]. In contrast, of key step in present scheme are the manipulation of spatially inhomogeneous polarizations in both arms and further the revelation of their quantum entangled nature. It is assisted by the combination of a detector (D), a single-mode fiber (SMF), a horizontal polarizer, a half-wave plate and one *q*-plate. In the transmitted arm, the fast axis of half-wave plate is set at angle $\theta/2$ and inversely rotates the horizontal polarization by $\theta$ with respect to the *q*-plate. As the SMF exclusively selects the fundamental Gaussian mode of $m = 0$, then the combination, as indicated by Eq. (2), defines a detected state,

$$|\psi\rangle_t = \left(e^{-i\theta}|R,+2q\rangle_t + e^{i\theta}|L,-2q\rangle_t\right)/\sqrt{2}. \quad (5)$$

When the coincidence circuit clicks, the reflected photon is simultaneously collapsed to,

$$\begin{aligned}|\psi\rangle_r &= \langle\psi_t|\Psi_{hyper}\rangle \\ &= \left(e^{i\theta}|R,-2q\rangle_r + e^{-i\theta}|L,+2q\rangle_r\right)/\sqrt{2},\end{aligned} \quad (6)$$

which implies if we preset the $q$-plate in the reflected arm to satisfy $q_r = -q_t$ and $\theta_r = -\theta_t$, then we obtain a unit coincidence rate, since $|\psi\rangle_t$ and $|\psi\rangle_r$ form a pair of conjugate states. In the context of violation of the original Bell inequality, it is however essential to investigate the rotational effect of measuring apparatus on the coincidence rates [2]. Hence, we impart a simultaneous rotation by $\beta_t$ and $\beta_r$ to both the $q$-plate and the polarizer in the transmitted and reflected arms, respectively,

$$\hat{R}(\beta_t)|\psi\rangle_t = \frac{1}{\sqrt{2}}\left(e^{-i(2q\beta_t+\theta)}|R,+2q\rangle_t + e^{i(2q\beta_t+\theta)}|L,-2q\rangle_t\right), \tag{7}$$

$$\hat{R}(\beta_r)|\psi\rangle_r = \frac{1}{\sqrt{2}}\left(e^{i(2q\beta_r+\theta)}|R,-2q\rangle_r + e^{-i(2q\beta_r+\theta)}|L,+2q\rangle_r\right), \tag{8}$$

Subsequently, the yielding coincidence rate is given by,

$$C(\beta_t,\beta_r) = \left|\langle\Psi_{hyper}|\hat{R}(\beta_t)|\psi\rangle_t \hat{R}(\beta_r)|\psi\rangle_r\right|^2 = \cos^2[2q(\beta_t-\beta_r)]. \tag{9}$$

The sinusoidal behavior depending only on the difference, $\beta_t - \beta_r$, predicts an entanglement residing in a two-dimensional subspace. The physics underlying is that the two-photon state post-selected by the cascading $q$-plate and half-wave plate in both arms is exactly a maximally entangled Bell state,

$$|\Psi\rangle_{Bell} = (|\xi\rangle_t|\eta\rangle_r + |\eta\rangle_t|\xi\rangle_r)/\sqrt{2}, \tag{10}$$

where $|\xi\rangle = |R,+2q\rangle$ and $|\eta\rangle = |L,-2q\rangle$ are the *composite* spin-orbit bases. Following the idea of Langford *et al*. [14, 16], equation (10) is just the so-called spin-orbit hypoentangled states residing in a two-dimensional subspace. Our present post-selection by cascading *q*-plates and half-wave plates is quite analogous to that of an OAM Bell state post-selected by angular sectors [4] or vortex links entanglement post-selected by computer-controlled holograms [5]. Along the same line, we further test a symmetry version of the Clauser-Horne-Shimony-Holt (CHSH) Bell inequality [31] to verify the quantum entangled nature. The CHSH-Bell inequality places constraints on the Bell parameter, $S = E(\beta_t, \beta_r) - E(\beta_t, \beta_r') + E(\beta_t', \beta_r) + E(\beta_t', \beta_r')$, with $|S| \leq 2$. In our scheme, $E(\beta_t, \beta_r)$ can be calculated from Eq. (10) and expressed as

$$E(\beta_t, \beta_r) = \frac{C(\beta_t, \beta_r) + C(\bar{\beta}_t, \bar{\beta}_r) - C(\beta_t, \bar{\beta}_r) - C(\bar{\beta}_t, \beta_r)}{C(\beta_t, \beta_r) + C(\bar{\beta}_t, \bar{\beta}_r) + C(\beta_t, \bar{\beta}_r) + C(\bar{\beta}_t, \beta_r)}, \tag{11}$$

where $\bar{\beta} = \beta + \pi/4q$. It can be deduced from Fig. 3 that by properly choosing $\beta_t = 0$, $\beta_r = \pi/16q$, $\beta_t' = \pi/8q$ and $\beta_r' = 3\pi/16q$, we can, in theory, maximally violate the CHSH-Bell inequality with $S = 2\sqrt{2}$. It thus suggests that present spin-orbit hypoentanglement of Eq. (10) describes quantum correlations between spatially inhomogeneous polarizations of SPDC photon pairs. Of further interest is that, quantum correlations, as indicated by Eqs. (5) and (6), only exist for those cylindrical polarizations described by a pair of conjugated states, such as those in Figs. 1(c) and 1(d) or in Figs. 1(e) and 1(f), where the instantaneous electric fields of linear polarization in a photon pair always oscillate in the opposite directions. While in Figs. 1(a) and 1(b) they are completely not quantum correlated.

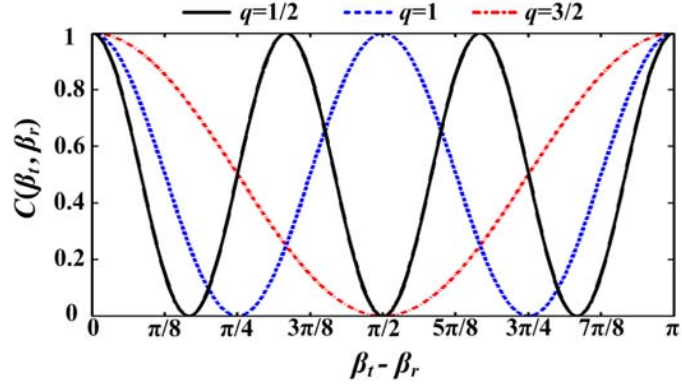

Fig. 3 (Color online) The predicted coincidence rate for accessing the violation of the CHSH-Bell inequality.

## 4. CONCLUSIVE REMARKS

We have theoretically demonstrated the violation of a Bell inequality for two-dimensional spin-orbit hypoentanglement and therefore revealed the entangled nature of cylindrical polarization symmetry of SPDC photon pairs. It is noted that, starting with a pure polarization entanglement (without OAM entanglement), one could also create the state of Eq. (10) by inserting a suitable $q$-plate in each arm [21]. However, this will require additional $q$-plates and half-wave plates to perform the state analysis. In contrast, present scheme is based on an initial spin-orbit hyperentangled state, and therefore can facilitate the experimental setup obviously. Besides, it was known that a radially polarized light can create a sharper focus [32] such that present manipulation of cylindrical polarizations of entangled photon pairs may show potential in high-contrast ghost imaging [33]. Besides, there is increasing interest of spatially inhomogeneous polarizations in quantum optics [34]. For example, it was recently demonstrated that a class of optical fibers are able to transport the cylindrical polarizations over long lengths with unprecedented stability [35] such that present work promises the fiber transport of spatially entangled photons, and therefore leading to a better security and resilience in quantum cryptography [36].

## ACKNOWLEDGEMENTS

The authors gratefully acknowledge the support from the Natural Science Foundation of China (Grant No. 11104233), the Natural Science Foundation of Fujian Province of China (Grant No. 2011J05010) and the Fundamental Research Funds for the Central Universities (Grant No. 2011121043).